%% file: IEEE_IPCCC_2019_CameraReady.tex
\documentclass[conference]{IEEEtran}
\IEEEoverridecommandlockouts
\usepackage{cite}
\usepackage{amsmath,amssymb,amsfonts}
\usepackage{algorithmic}
\usepackage{textcomp}
\usepackage{xcolor}
\usepackage{multirow}
\usepackage{subfigure}
\usepackage[pdftex]{graphicx}

\makeatletter
\def\ps@IEEEtitlepagestyle{
  \def\@oddfoot{\mycopyrightnotice}
  \def\@evenfoot{}
}
\def\mycopyrightnotice{
  {\footnotesize 978-1-7281-1025-7/19/\$31.00~\copyright~2019 IEEE\hfill} 
  \gdef\mycopyrightnotice{}
}

\def\BibTeX{{\rm B\kern-.05em{\sc i\kern-.025em b}\kern-.08em
    T\kern-.1667em\lower.7ex\hbox{E}\kern-.125emX}}
    
\newcounter{RZNumberOfComments}
\stepcounter{RZNumberOfComments}

\begin{document}
\title{Uncovering Flaming Events on News Media in Social Media}
\author{\IEEEauthorblockN{Praboda Rajapaksha\IEEEauthorrefmark{1}\IEEEauthorrefmark{2}, Reza Farahbakhsh\IEEEauthorrefmark{1}, No\"{e}l Crespi\IEEEauthorrefmark{1}, Bruno Defude\IEEEauthorrefmark{1}}
\IEEEauthorblockA{\IEEEauthorrefmark{1}\textit{Institut Polytechnique de Paris, Telecom SudParis, CNRS Lab UMR5157, Evry, France.}\\.
\IEEEauthorblockA{\IEEEauthorrefmark{2}\textit{Uva Wellassa University, 90000, Badulla, Sri Lanka.}}
\{praboda.rajapaksha, reza.farahbakhsh, noel.crespi, bruno.defude\}@telecom-sudparis.eu}
}
\maketitle

\input{1_Abstract}
\begin{IEEEkeywords}
Flaming detection, Sentiment analysis, deep neural networks, social media, Facebook, News media, FastText, Word2Vec.
\end{IEEEkeywords}
\input{2_Introduction}
\input{3_Related_work}	
\input{4_Methodology}
\input{5_Analysis}
\input{6_Conclusion}

\bibliographystyle{IEEEtran}  
\bibliography{citation}

\end{document}

%% file: 1_Abstract.tex
\begin{abstract}
Social networking sites (SNSs) facilitate the sharing of ideas and information through different types of feedback including publishing posts, leaving comments and other type of reactions. However, some comments or feedback on SNSs are inconsiderate and offensive, and sometimes this type of feedback has a very negative  effect on a target user. The phenomenon known as flaming goes hand-in-hand with this type of posting that can trigger almost instantly on SNSs. Most popular users such as celebrities, politicians and news media are the major victims of the flaming behaviors and so detecting these types of events will be useful and appreciated. Flaming event can be monitored and identified by analyzing negative comments received on a post. Thus, our main objective of this study is to identify a way to detect flaming events in SNS using a sentiment prediction method. We use a deep Neural Network (NN) model that can identity sentiments of variable length sentences and classifies the sentiment of SNSs content (both comments and posts) to discover flaming events. Our deep NN model uses Word2Vec and FastText word embedding methods as its training to explore which method is the most appropriate. The labeled dataset for training the deep NN is generated using an enhanced lexicon based approach. Our deep NN model classifies the sentiment of a sentence into five classes: Very Positive, Positive, Neutral, Negative and Very Negative. To detect flaming incidents, we focus only on the comments classified into the Negative and Very Negative classes. As a use-case, we try to explore the flaming phenomena in the news media domain and therefore we focused on news items posted by three popular news media on Facebook (BBCNews, CNN and FoxNews) to train and test the model. The experimental results show that flaming events can be detected with our proposed approach, and we explored main characteristics that trigger a flaming event and topics discussed in the flaming posts.
\end{abstract}

%% file: 2_Introduction.tex
\section{Introduction}
 In modern Internet parlance, cyberbullying has become increasingly common, especially in the Internet communities and SNSs such as Facebook, Twitter, Instagram and YouTube. In this era, several new phenomena have appeared, leading to important debates and discussions and among them one hot topic is `Flamings'. The Flaming can be considered as a serious issue on SNSs where many users express disagreements, insults or offensive words in the form of comments on a forum, blog or chat room intended to inflame emotions and sensibilities of others. These comments do not contribute any useful content to the discussion groups and instead attempt to wound another person socially and psychologically. These comments might be posted by genuine users and fraud or spam generated content \cite{44}.
Two examples of flaming include 1) `Delete your account': a Clinton-Trump Twitter flame  war\footnote{https://www.sbs.com.au/news/delete-your-account-clinton-trump-in-twitter-flame-war} and 2) flame war between Donald Trump and Pope Francis on the pope’s calling Trump `disgraceful' for his immigration recommendations\footnote{https://www.lifewire.com/what-is-flaming-2483253}. As a result of this high-level visibility, flaming has became an interesting topic among social researchers  as they seek to understand the phenomena and explore the impact of these types of activities on targets.  
 
One use-case we target in this study is news media in Facebook as many American adults consume news on social media and majority of them are commonly use Facebook \cite{19}. In addition we can observe that the number of fans in news media Facebook pages\footnote{\label{note1}https://fanpagelist.com/category/news/} are considerably higher similar to other categories. 
News media tend to publish news items of interest to more diversified and varied readers, and therefore many news readers interact with news items daily via commenting, sharing and reacting. A set of news media is relying on the content of most popular news media as content consumers \cite{42},\cite{43} and so, it is important to understand these flaming types of events in the news media domain. 
Thus, our main objective is to explore news items' flaming events in terms of negative feedback with insults and other offensive words.
The existence of flamings on news items may reduce number of followers, or sometimes these types of posts can go viral and increase the number of followers. Therefore detecting flamings and identifying the topics that the community most strongly disagree with is a useful conception for news media.

Sentiment polarity prediction is one of the main ways of detecting flaming events in SNS \cite{10}, \cite{11}. 
Many advancements have been made to the sentiment classification methods to date. However, these methods are domain-specific and therefore their results are strongly biased on the words used in that domain. 
As a result, we build a word embedding-based multiclass (5-classes) sentiment classification deep NN-based approach by focusing on Facebook news items and variable length user feedbacks that can be modified and applied in any other social media category other than news.
In addition, we use an improved lexicon-based sentiment classification method to generate a true labeling list with which to train the deep NN model. 
The flaming effect analyses will be done based on the comments classified as Negative and Very Negative. A flaming event takes place when many users give negative feedback, and so a post with a large number of negative comments received within a short time will possibly a flaming event. We will also explore what types of topics were mainly affected by flamings that are published by BBCNews, CNN and FoxNews.

This paper offers the following contributions:\\
1) a word embedding-based sentiment prediction deep NN model focused on Facebook comments.\\
2) an exploration of which word embedding method (Word2Vec or FastText) works better on Facebook comments.\\
3) identification of flaming posts on BBCNews, CNN and FoxNews Facebook pages published in February 2018. \\
4) identification of flaming posts' associated topics.



%% file: 3_Related_work.tex
\section{Literature Survey}
\label{lbl:liturature_survey}


Researchers have investigated flamings on YouTube \cite{9}, email threads \cite{24} and comments received on Twitter and Facebook  \cite{10}, \cite{11}, news sites and news channels \cite{23}. These works show that flaming events can be appeared in any online platforms especially in the SNSs as any user can comment on a public content.
Sentiment polarity prediction is one of the main ways of detecting flaming events in SNSs \cite{10}, \cite{11}.

 Sentiment analysis of social media content has become more and more popular, as it can be used for mining opinions on services, products, companies, etc. and these models can be implemented as supervised, unsupervised or semi-supervised  \cite{29} methods.
However, with the increase of user-generated content in SNSs it is not effective to apply lexicon-based unsupervised methods, and therefore supervised methods can be automated to detect polarity. 
Apart from that, analyses on sentiment classification are aspect-based and domain-specific. Therefore, model needs to train with domain specific data to achieve better accuracy when building a sentiment prediction model. Recently, it has been widely acknowledged that deep learning-based representation models have achieved great success in text sentiment classification tasks compared to traditional machine learning models \cite{32}.
Furthermore, in recent works,  word embedding-based method are applied for sentiment classification. A few have used Word2Vec embeddings \cite{26}, \cite{27}.
Deep learning has emerged as a powerful machine learning technique and is also popularly used in sentiment analysis in recent years \cite{38}, \cite{28}.
Wang et al. proposed a CNN-RNN architecture \cite{37} to analyze the sentiment of short text while some other studies tried apply methods based on CNN \cite{40} and RNN \cite{35}. Their experimental results shown that the proposed method outperforms lexicon-based, regression-based, and obtained an obvious improvement upon the state-of-the-art. In addition, many research works considered only the binary sentiment classification and few studies have used multi-class sentiment classification producing promising results \cite{41}. 

Hence, in this work we try to build a deep NN model combining CNN and RNN to classify sentiment of variable length text using word embeddings that can be generalized as a semi-supervised model to be adapted to any domain specific dataset. Also, the model will consider multi-class labels to train and experiment with a real datasets. 

%% file: 4_Methodology.tex
	
\section{Experiments}

\label{lbl:methodology}
News media flaming events can be detected by identifying negative comments received on shared news items in SNSs. Therefore, sentiment prediction models can be adapted to cluster senses of the user feedbacks in order to explore the existence of flaming events in news media in SNS using a rule based technique.
Many previous sentiment analysis works have used Twitter as the SNS to analyze and build their sentiment prediction models. As Facebook employ different content properties such as permitting to share variable length posts and comments, unlike in Twitter, we build our methodology by proposing a sentiment prediction model focusing on news items shared on Facebook. Most efficient sentiment analysis algorithms are supervised learning which requires sufficient amount of training data. Hence, we introduce a deep-NN based supervised sentiment prediction model that require considerable number of true labels to train. The labeled dataset is generated using an unsupervised model which is an improved unsupervised sentiment classification model. In the following sections we explain our sentiment classification deep NN-based approach in detail.

\subsection{Dataset description and pre-processing}
 In this study, we try to explore existence of the flaming events in news media in Facebook as news media plays a major role to promote and distribute their news items in Facebook  \cite{19}.
Therefore, this study targets on top three news media (BBCNews, CNN and FoxNews) in Facebook having the highest number of fans as a usecase to explore flamings. 

We implemented a crawler to collect ramdom 300 public posts and related comments of BBCNews, CNN and FoxNews in February 2018 using Facebook Graph API. To respect the ethical aspects, we collected only the texts of public posts, comments and timestamps from their Facebook accounts and do not collected neither sensitive data nor personal data. The analyses are mainly based on this dataset and a brief description of the dataset is presented in Table \ref{tab:dataset}. The training and experimental datasets are separated after pre-processing the original news items and user feedback.

\begin{table}[]
\centering
\caption{Statistics of public posts obtained from BBC, CNN, and FoxNews Facebook pages during February 2018. (PP  represents the abbreviation for Pre-Processing)}
\scriptsize
\label{tab:dataset}
\begin{tabular}{|l|l|l|l|l|l|}
\hline
\multirow{2}{*}{\textbf{\begin{tabular}[c]{@{}l@{}}News\\ media\end{tabular}}} & \multirow{2}{*}{\textbf{\#Fans}} & \multicolumn{4}{c|}{\textbf{Number of comments of}}                                                                                                                                                                                                                                               \\ \cline{3-6} 
                                                                               &                                  & \textbf{\begin{tabular}[c]{@{}l@{}}300 posts\\ Before PP\end{tabular}} & \textbf{\begin{tabular}[c]{@{}l@{}}300 posts\\ After PP\end{tabular}} & \textbf{\begin{tabular}[c]{@{}l@{}}100 PPposts\\ Training\end{tabular}} & \textbf{\begin{tabular}[c]{@{}l@{}}200 PPposts\\ Experiment\end{tabular}} \\ \hline
\textbf{BBCNews}                                                               & 46.2M                            & 398,453                                                                & 312,881                                                               & 107,874                                                              & 205,007                                                                   \\ \hline
\textbf{CNN}                                                                   & 29.9M                            & 595,268                                                                & 312,881                                                               & 217,386                                                              & 288,606                                                                   \\ \hline
\textbf{FoxNews}                                                               & 16.3M                            & 1,162,734                                                              & 312,881                                                               & 280,342                                                              & 773,439                                                                   \\ \hline
\end{tabular}
\end{table}


We applied basic preprocessing techniques such as replace URL with a space, remove user mentions, hashtags, retweets, special characters and multiple spaces, removed multiple letters from a word ('haaappy' will be replaced by 'happy') ,merge characters of a word written with spaces or dots in between each character ('h a p p y', 'h.a.p.p.y' will be reformatted to happy). As the objective of this work is to identify the sentiment of SNSs content, emojis impact a lot on the overall sentiment value. Therefore, we included all the positive and negative emojis in the text without ignoring.
We also performed stemming on each word in the sentence to generate the stem. Lemmatizaiton is not applied as lemmatizers have to search through a corpus while stemmers do basic string operations and thus is faster than lemmatizers. Also, stemming can work with unknown words while lemmatizer do not recognize them. We feed to the neutral network separately the datasets after being lemmatized and stemmed, and the accuracy of the model was very low with the lemmatized dataset. We do not remove stopwords in sentences while pre-processing.

Considerable amount of comments have been removed after preprocessing and the main reason was that the majority of those comments were URLs. We can observe from the Table \ref{tab:dataset} that news items published by FoxNews has received a higher number of comments, while having less number of fans, than BBCNews and CNN. We use comments of 100 pre-processed posts from each news media as a training dataset to train the NN model and comments of the remaining 200 news posts from each news media as the experimental dataset to explore flaming events.

\begin{figure}[t]
\centering
\includegraphics[width=3.2in]{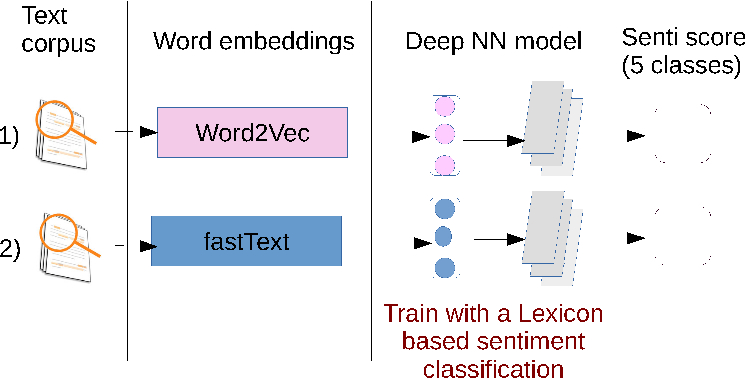}
\caption{The architecture of the multiclass sentiment prediction using word embeddings}
\label{fig:main_architecture}
\vspace{-0.4cm}
\end{figure}

\subsection{Proposed sentiment analysis architecture}
\label{sec:sentiment_architecture}
The architecture of our variable length text sentiment prediction model is shown in Figure \ref{fig:main_architecture}. 
At first, as depicted in Figure \ref{fig:main_architecture}, we build a lexicon-based approach to generate the training dataset for our deep NN model. The technique we used in the lexicon-based approach is explained in detail in Section \ref{true_labels}. 
Many previous studies have tried to employ word-embedding methods for machine leaning or deep learning-based sentiment classification \cite{28},\cite{26},\cite{27}. 
In our prediction model we experiment with two different word embedding methods: Word2Vec and FastText. 
Word2Vec is widely used for both shallow and deep neural networks and was developed by Google in 2013 \cite{4}. In the Word2Vec model, words with similar meaning are mapped to a nearby vector space, thereby making it simple to explore semantically similar words. As the Word2Vec model is based on a predefined dictionary, one shortcoming of this representation is that rare words may not be mapped with other vectors. However, FastText was developed by Facebook \cite{5} in 2016 using the n-gram representation of each word in a sentence and therefore helps to detect misspelled words and slang across different languages. 
In the experiments we will explore which word embedding method is more suitable for sentiment classification of the Facebook content.

The deep NN model shown in Figure \ref{fig:main_architecture} uses the best performed word embedding method to predict the sentiment of a sentence based on a multiclass (5 classes) classification technique, where five different classes can be defined as;  Very Positive(4), Positive(3), Neutral(2), Negative(1) and Very negative(0). The accuracy of the model can be improved by re-training the model with experimental datasets by increasing the labeled dataset occasionally as a semi-supervised approach.

\makeatletter
\newcommand\footnoteref[1]{\protected@xdef\@thefnmark{\ref{#1}}\@footnotemark}
\makeatother

\subsection{Methodology for generating true lables}
\label{true_labels}

The deep NN sentiment prediction model need to use a set of true labels to train the model. We introduce an improved lexicon-based approach to identify the sentiment of Facebook content and then use this model to build the training corpus for a deep NN model to automatically classify the sentiments. In this approach we can customize sentiment labels to be 2-class (positive, negative), 3-class (positive, neutral, negative), or 5-class(very positive, positive, neutral, negative, very negative) depending on the requirement.
This section explains this unsupervised lexicons-based sentiment classification approach.
 
The consideration of negators, adverbs and model verbs associated with words is important for sentiment classification. 
Therefore it is worth to identify important sentiment of words and word phrases separately including various negators, models, and degree adverbs, as well as their combinations. 
Kiritchenko et al. \cite{7} created a lexicon dataset, Sentiment Composition Lexicon for Negators, Models, and Degree Adverbs (SCL-NMA) and it includes 3207 phrases with related sentiment scores where each phrase contains at most 4 words. Sentiment value of a phrase ranges from -1 to 1 where phrases with negative meanings have assigned negative values and phrases with positive meanings have assigned positive values. The higher the meaning of a phrase is positive the more it is closer to 1 and similarly, sentences having very negative meanings have assigned score very closer to -1. 

We used an statistical based approach to identify the polarity of a sentence in English. A sentence may have one or many lexicons found in the SCL-NMA dataset and also both negative and positive lexicons. Hence, we explored all the lexicons in the sentence to classify its overall sentiment. This approach is very useful to identify the sentiment of variable length sentences and following equation shows the prediction of sentiment of a new sentence. Assume that SCL-NMA dataset consists of lexicons L1, L2,...,Lx. 
\begin{equation} \label{eq:1}
SentiScore = \dfrac{[\sum_{n=1}^{N} Ln ]+ [C + S + E]}{[N + C + S + E]} 
\end{equation}
Where N is the number of lexicons in the sentence. Letters C, S, and E are constant and set to be 1, -1 or 0 depending on the lexicon polarity. 
These constant values are added to the SentiScore formula based on the existence of different properties of the lexicons in a sentence as explained  below. 
 
The overall sentiment of a sentence is increased when it has lexicons (found in the SCL-NMA dataset) with capital letters. Therefore, we identify those lexicons in capital letters and use these details in Equation \ref{eq:1}, the letter C, to evaluate the overall sentiment score.
\begin{equation} \label{eq:2}
C = \sum capital\_lexicons
\end{equation}

where \textit{capital\_lexicons=1}, if the lexicons' in capital letters has a positive sentiment value, \textit{capital\_lexicons=-1}, if the lexicons' in capital letters has a negative sentiment value and \textit{capital\_lexicons=0}, if no capitalized lexicons in the sentence. 

Emojis in a sentence are affecting more on the sentiment score. Thus, E in Equation \ref{eq:1} refers to all emojis in the sentence those that helps to increase the strength of the sentiment. 
\begin{equation} \label{eq:2}
E = \sum emojis
\end{equation}
where \textit{emojis=1} for a positive emoji,  \textit{emojis=-1} for a negative emoji and  \textit{emojis=0} if no emojis presence in a sentence. 

People use exclamation marks to stress words and this is an important property when evaluating sentiments. Thus, we observe the presence of exclamation marks in a sentence and this feature is presented in Equation \ref{eq:1} as S.
\begin{equation} \label{eq:2}
S = exclamation\_mark
\end{equation}
where \textit{exclamation\_mark=1} if a positive lexicon is attached to exclamation mark, \textit{exclamation\_mark=-1} if a negative lexicon is attached to exclamation mark, and \textit{exclamation\_mark=0} if no any exclamation mark is attached to lexicons. 


\begin{table}[]
\caption{Confusion matrix of our lexicon-based sentiment classification method (macro precision-60.66\%, macro recall-62.01\%, F1 score - 61.31\%)}
\vspace{-0.4cm}

\label{tbl:lexicon}
\center
\begin{tabular}{cl|l|l|l|l|}
\cline{3-6}
\multicolumn{1}{l}{}                                   &              & \multicolumn{4}{c|}{\textbf{Predicted}}                         \\ \hline
\multicolumn{1}{|c|}{\multirow{4}{*}{\textbf{Actual}}} &              & \textbf{Pos} & \textbf{Neg} & \textbf{Neu} & \textbf{Precision} \\ \cline{2-6} 
\multicolumn{1}{|c|}{}                                 & \textbf{Pos} & 128          & 19           & 35           & 70.33\%            \\ \cline{2-6} 
\multicolumn{1}{|c|}{}                                 & \textbf{Neg} & 57           & 83           & 37           & 46.89\%            \\ \cline{2-6} 
\multicolumn{1}{|c|}{}                                 & \textbf{Neu} & 37           & 12           & 90           & 64.75\%            \\ \hline
\multicolumn{1}{|l|}{\textbf{Recall}}                  &              & 57.66\%      & 72.81\%      & 55.56\%      &                    \\ \hline
\end{tabular}
\end{table}

\begin{table}[]
\caption{Confusion matrix of Vadar sentiment classification (macro precision-37.85\%, macro recall-51.41\%, F1 score - 43.59\%)}
\vspace{-0.4cm}

\label{tbl:vadar}
\center
\begin{tabular}{cl|l|l|l|l|}
\cline{3-6}
\multicolumn{1}{l}{}                                   &              & \multicolumn{4}{c|}{\textbf{Predicted}}                         \\ \hline
\multicolumn{1}{|c|}{\multirow{4}{*}{\textbf{Actual}}} &              & \textbf{Pos} & \textbf{Neg} & \textbf{Neu} & \textbf{Precision} \\ \cline{2-6} 
\multicolumn{1}{|c|}{}                                 & \textbf{Pos} & 24           & 12           & 156          & 12.5\%             \\ \cline{2-6} 
\multicolumn{1}{|c|}{}                                 & \textbf{Neg} & 0            & 15           & 162          & 8.46\%             \\ \cline{2-6} 
\multicolumn{1}{|c|}{}                                 & \textbf{Neu} & 1            & 1            & 137          & 92.57\%            \\ \hline
\multicolumn{1}{|l|}{\textbf{Recall}}                  &              & 70.56\%      & 53.57\%      & 30.11\%      &                    \\ \hline
\end{tabular}
\end{table}

The algorithm consider n-grams as the features, where n ranges from 1 to 3 and generate features separately for lexicon dataset and sentences that need to identify its sentiments.  
In summary, Equation \ref{eq:1} helps to identify the sentiment of any sentence written in English regardless of the length of the sentence. This approach helps to generate a labeling dataset from user feedback received on news items shared on Facebook in order to train our deep NN model .

\subsection{Comparing unsupervised true lable generation method with a baseline method}

VADAR lexicon and rule-based sentiment analysis tool \cite{30}, which is widely used for classifying social media content, is considered as the baseline method to compare our unsupervised true label generation model.
For a given sentence, VADAR returns its sense from 3 classes: positive, negative or neutral. Therefore, in order to compare our algorithm with VADAR, our model is customized to classify sentence sentiment in to the same 3 classes.

We use a manually annotated dataset having 498 sentences with its sentiment values: 177 negative tweets, 182 positive tweets and 139 neutral tweets published by the Stanford University \cite{31}.
This dataset is applied on both VADAR and our unsupervised sentiment classification technique and their confusion matrices are shown in Table \ref{tbl:vadar} and Table \ref{tbl:lexicon}, respectively.
The precision, recall and F1 score parameters of our approach is much higher than VADAR for the classification of positive and negative sentiments as detailed in Table \ref{tbl:vadar} nd Table \ref{tbl:lexicon}. 
The results is proven that our lexicon based approach can classify the sentiment label of a sentence much better than VADAR. However, we can still improve the accuracy of our model by increasing the lexicon dataset with additional lexicons. 

This approach is unique as our training dataset for the deep NN can be generated automatically, as opposed to having manual annotate Facebook posts and comments. Moreover, this model is introduced as a multiclass sentiment classification approach rather than classifying only into 2 or 3 classes. In this model, we assigned a sentiment label for each sentence from five classes as follows. 
\normalsize

\footnotesize
\begin{equation} 
Prediction=
\begin{cases}
VeryPositive, IF (sentiScore => 0.5)\\
Positive, IF (0.5 > sentiScore > 0)\\
Neutral, IF (sentiScore = 0)\\
Negative, IF (0 > sentiScore > -0.5)\\
VeryNegative, IF (sentiScore <= -0.5)\\
\end{cases}
\end{equation}
\normalsize
The baseline value for distinguishing Very Positive from Positive and Very Negative from Negative classes is set to be 0.5 and -0.5 respectively. 
 \begin{figure}[t]
\centering
\includegraphics[width=3.2in]{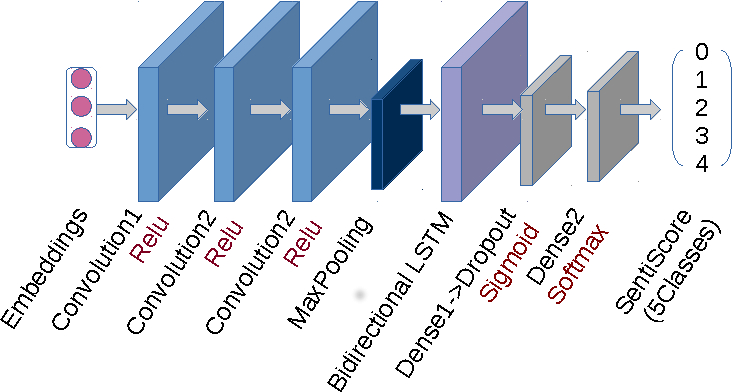}
\caption{The deep model for extracting sentiment scores}
\label{fig:deepNN_model}
\end{figure}

\begin{figure*}[t]
	\centering
	\subfigure[BBCNews (neu-41\%, pos-38\%, neg-21\%)]{
  \includegraphics[width=2.48in]{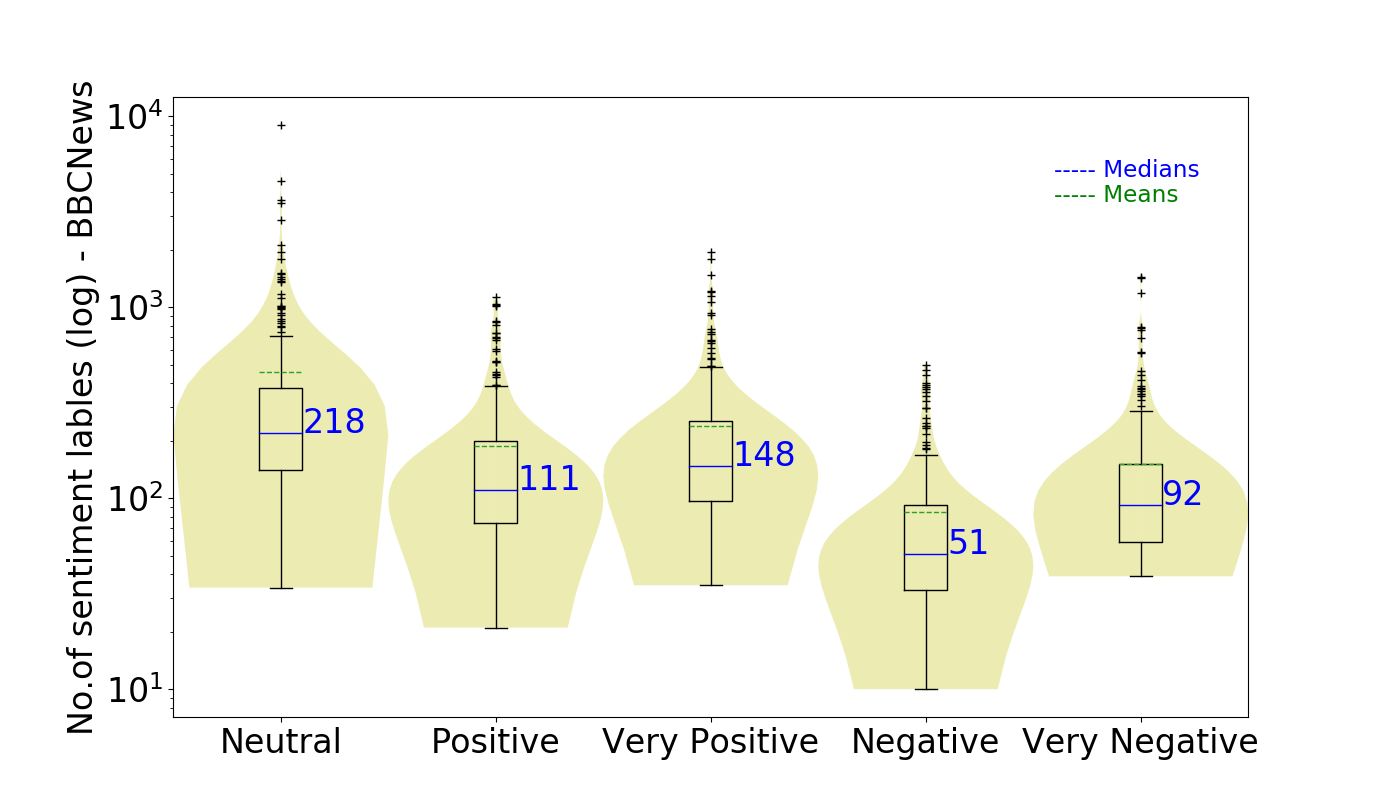}
\label{fig:post1}
	}   \hspace{-1cm}
	\subfigure[CNN (neu-39\%, pos-38\%, neg-23\%)]{
  \includegraphics[width=2.49in]{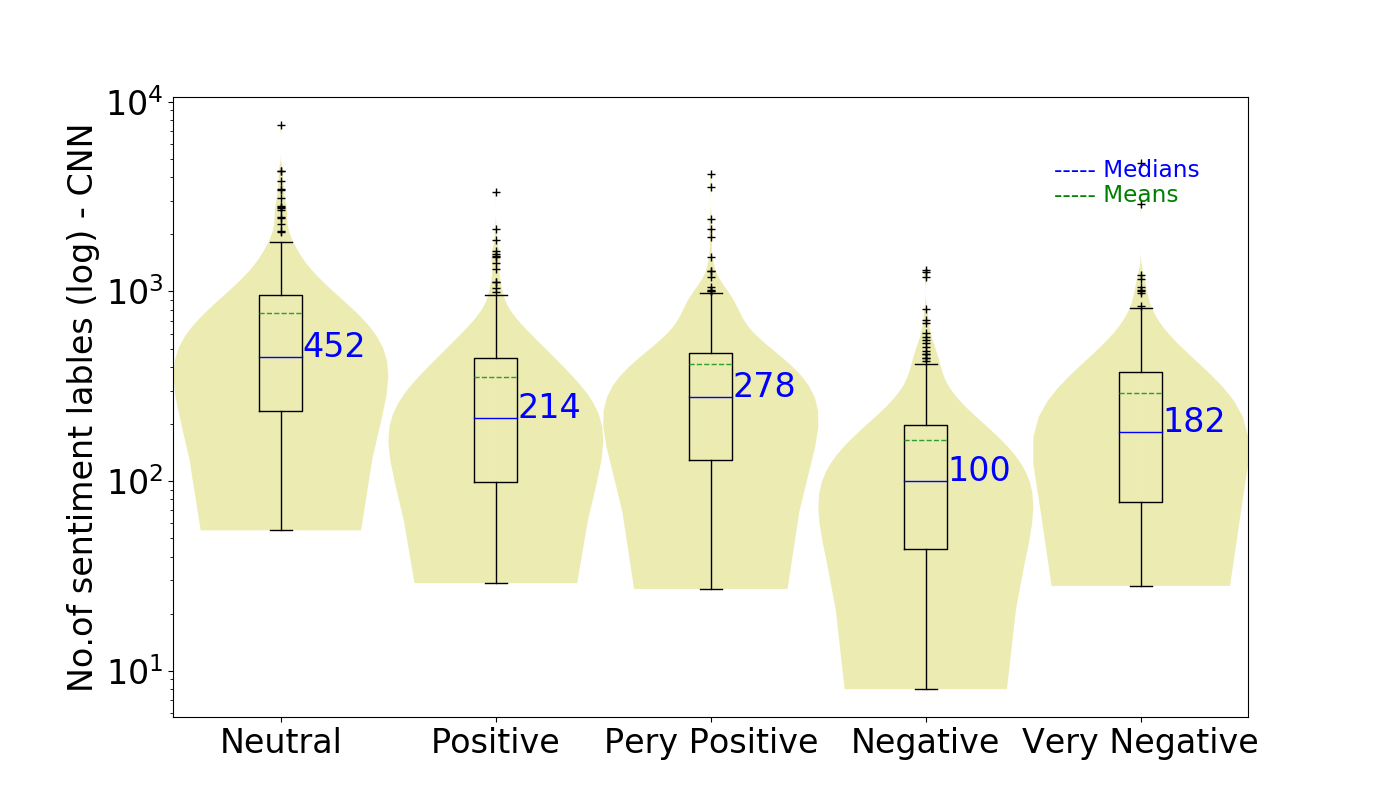}
\label{fig:post2}
	}   \hspace{-1cm}
	\subfigure[FoxNews (neu-40\%, pos-36\%, neg-24\%)]{
  \includegraphics[width=2.49in]{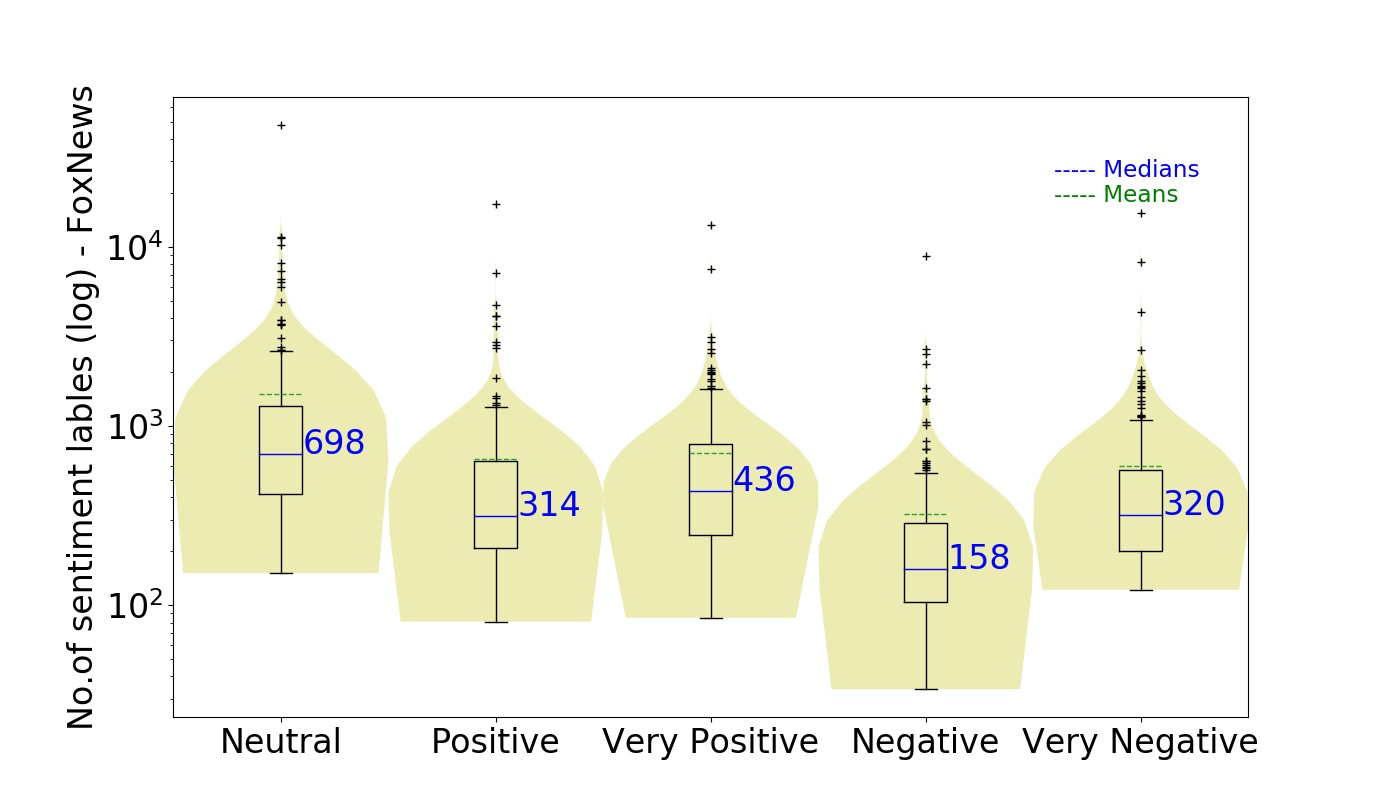}
\label{fig:post3}
	}
	\caption{Five class sentiment value distribution of 200 random posts shared by BBCNews, CNN and FoxNews during February 2018.}
	\label{fig:TotalSentiments}
		\vspace{-0.5cm}

\end{figure*}

\subsection{Methodology for an automatic sentiment prediction model}

Since deep learning-based sentiment classification models are better than the traditional machine learning models \cite{32} we try to adapt one deep learning-based model in this work to predict sentiments of Facebook content.
These deep models can learn text representation from original data that can capture relations between contexts words in a scalable manner. 

Recent research on text sentiment classification used convolution operations using n-gram features. However, the convolution neural network (CNN) completely ignores the sequence information of the text while paying attention to the local features of a sentence. On the other hand, long short-term memory (LSTM) networks are good at learning sequential correlation in the text by using an adaptive gating mechanism but lost the ability to learn local features of the text. Therefore, to involve both sequence correlation information and local features, it is important to explore an effective combination strategy that takes advantage of the CNN and the LSTM network. Hence, we experiment with several different architectural methods such as: 
i) CNN-LSTM-Word2Vec : pre-trained Word2Vec vectors, 
ii) CNN-LSTM-Word2Vec : custom Word2Vec model train with our own data, 
iii) CNN-LSTM-Fasttext : pre-trained Fasttext vectors, 
iv) CNN-LSTM-Fasttext : custom Fasttext model tran with our own data, and 
v) Adjusting the number of CNN layers and dropout layers, and maxpooling layer.

Models are trained separately with the custom Word2Vec and FastText embedding methods using 315K total number of comments (experimental dataset in Table \ref{tab:dataset}) with their sentiment labels generated from our lexicon-based method. The training dataset consist of 42.3\%-neutral, 20.5\%-very positive, 16.6\%-positive, 12.9\%-very negative and 7.7\%-negative labels. 
And also these models are trained with both pre-trained word embedding.
Most promising results with high accuracy is given to the architecture shown in Figure \ref{fig:deepNN_model} and achieved high accuracy with our own domain specific word embedding model rather than using pre-trained embedding. 

As shown in Figure \ref{fig:deepNN_model}, model was implemented with three CNN layers and a Bi-LSTM layer. The inputs of the NN model uses 1D convolutions.
One dropout layer was added within two dense layers (hidden layers) close to the output layer, and another dropout layer on the Bi-LSTM layer. 
The model compilation is done by adjusting three parameters: loss, optimizer and metrics. The Adam \cite{18} optimizer was used in our model to adjust the learning rate throughout the training and the learning rate was fixed to be very small (0.0001) leading to obtain more accurate results. The categorical\_crossentropy loss function was used in our implementation. 
This model uses three activation functions: Relu- for CNN layers, Softmax- for the output layer and Sigmoid- for the dense layer. 
The method of dealing with variable length sentences is that, at first we need to set a maximum number of token values (30 in this research work) and then if a comment has more than 30 tokens, we divided it into sub-sentences and evaluate the sentiment of each sub-sentence separately. Finally, the sentiment score for the entire sentence is the average sentiment scores of sub-sentences.



At first, deep NN model is trained for 40 epochs and evaluated the accuracy and validation loss for each iteration separately for Word2Vec and FastText. 
In the Word2Vec model, we analyzed that at the 13th epoch, the performance of the training dataset continue to decrease than the validation dataset, indicating an over-fitting. Therefore, the best-fit for this model is to training with 12 epoch as the model has good skill on both the training dataset and unseen test dataset. The accuracy of the model is identified as 85\%.
Similarly, when the FastText model used in the deep NN model, it shows the best fit after 15th epoch, in which training and validation loss is almost equal and with 78\% accuracy. In both scenarios, total number of parameters trained by the NN model is 3,050,053.

The results manifested that, Word2Vec model out performs the FastText model when applying for multiclass sentiment classification of Facebook content. As a result, we will use our generated models with Word2Vec embedding to explore the existence of any flaming events in news media in Facebook.


%% file: 5_Analysis.tex
\section{Analysis of flaming events in BBCNews, CNN and FoxNews in Facebook}
\label{lbl:analysis}

\begin{figure*}[t]
	\centering
	\subfigure[No.of N/VN comments received on posts per day]{
  \includegraphics[width=3.5in]{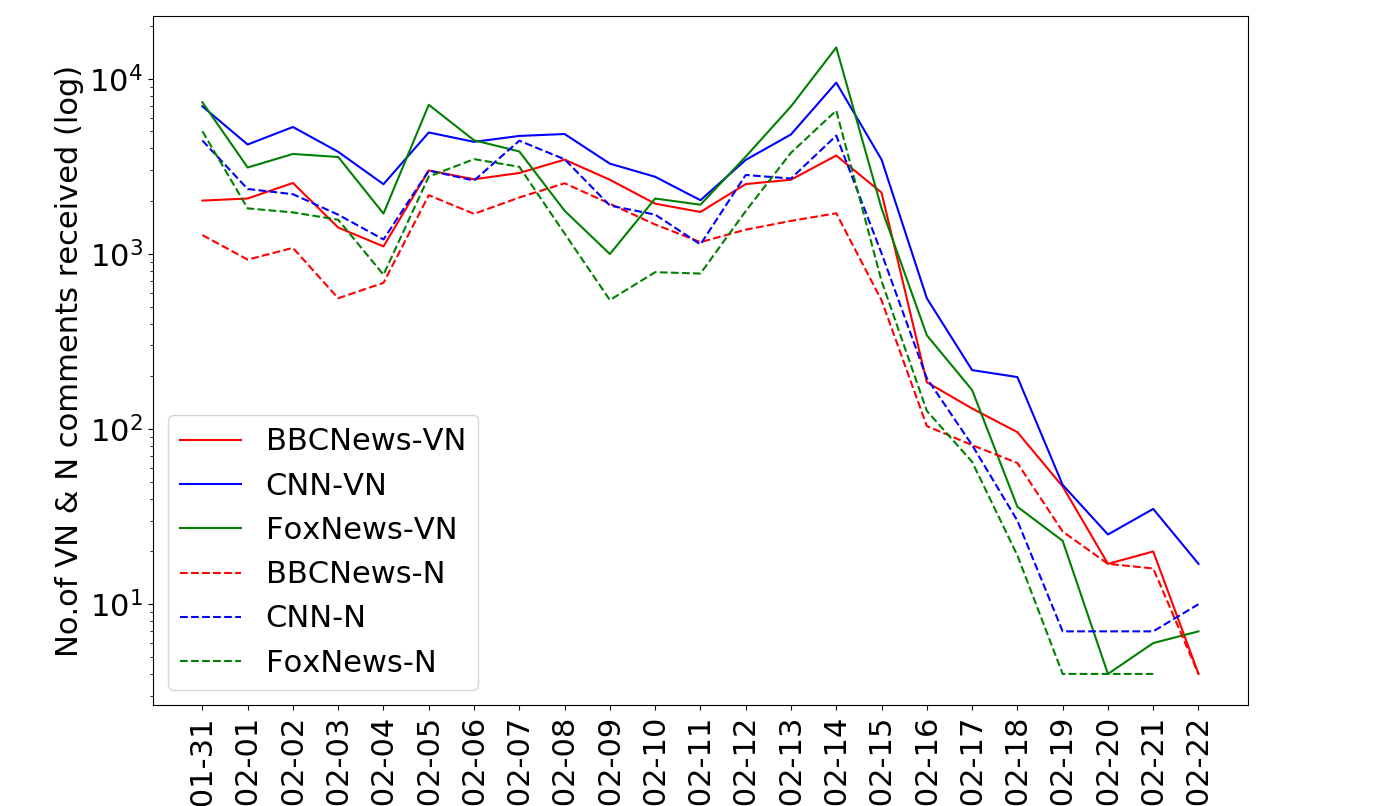}
\label{fig:timeseries-perday}
	}   \hspace{-1cm}
	\subfigure[No.of VN comments received on posts shared in 14th February 2018]{
  \includegraphics[width=3.5in]{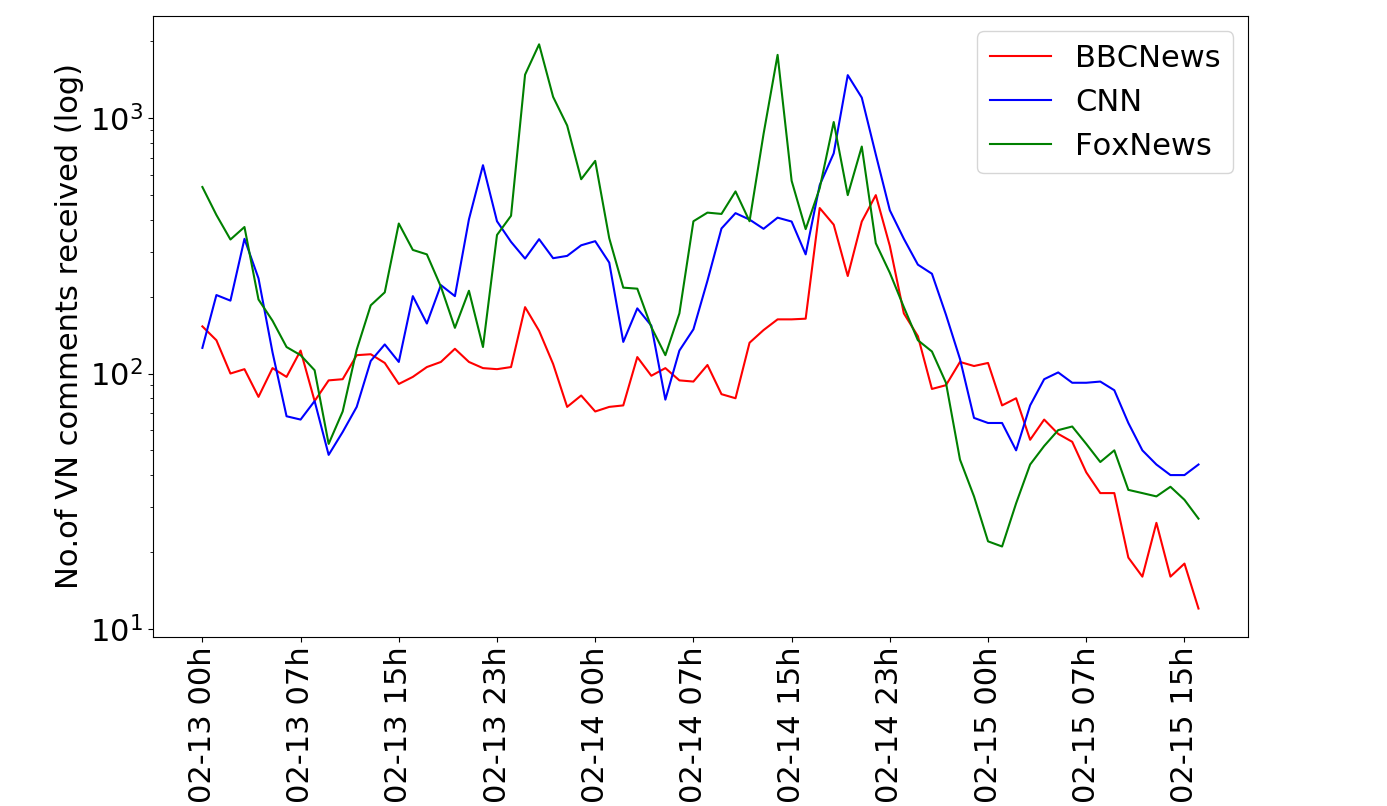}
\label{fig:timeseries-14thfeb}
	} 
	\caption{The distribution of Negative (N) and Very Negative (VN) comments received on posts shared by BBCNews, CNN and FoxNews in February 2018.}
		\vspace{-0.1cm}
	\label{fig:sentimentdistro_negative}
\end{figure*}

In social media, relatively large number of critical comments can be directed at individuals, companies, brands and etc. 
and this behavior is called as flaming. 
One way of exploring flaming behavior is to identify whether a post received higher number of very negative comments within a short time. This section explains, how the implemented deep NN sentiment prediction model can be used to identify the existence of flaming or similar kind of event in news media in Facebook. 


\subsection{Experimental dataset}
In order to analyze flaming events and their existence in news media in Facebook, we randomly selected comments from 200 posts (Table \ref{tab:dataset} and Figure \ref{fig:TotalSentiments}) shared by BBCNews, CNN and FoxNews in February 2018 as our experimental dataset. 
First, we try to explore sentiment labels for each comment in the experimental dataset using our deep NN model which is trained with the Word2Vec model as Word2Vec model given a high accuracy than the FastText model (Section \ref{lbl:methodology} ).

Figure \ref{fig:TotalSentiments} shows the statistics of the sentiment prediction of the comments belongs to 600 posts in our experimental dataset (200 posts from each BBCNews, CNN and FoxNews). As indicated in Figure \ref{fig:TotalSentiments}, number of Neutral comments received on these posts are much higher than the other classes. The second most number of sentiment predictions of comments are belongs to the Positive class, followed by Very Positive and Very Negative. The least number of predicted sentiment values are from the Negative class. In general, number of Neutral comments received on the posts are always higher than the total number of positive comments (Very Positive + Positive) and total number of negative comments (Very Negative + Negative).
We observed that the text classified into a Neutral class have a large number of stop-words, names, URLs, and single words without indication of any feelings. 

\begin{figure*}[t]
	\centering
	\subfigure[BBCNews]{
  \includegraphics[width=2.4in]{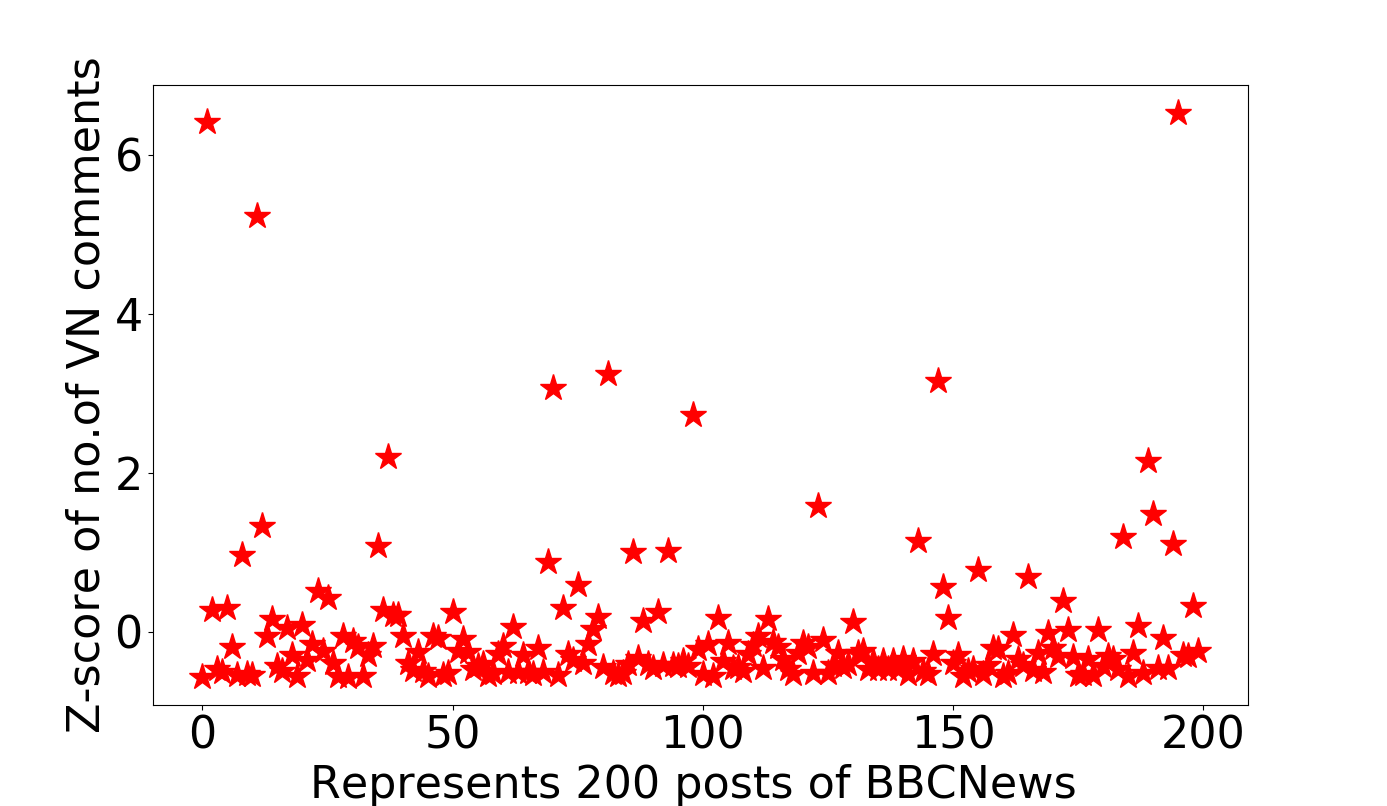}
\label{fig:z-score-bbc}
	}   \hspace{-0.8cm}
	\subfigure[CNN]{
  \includegraphics[width=2.4in]{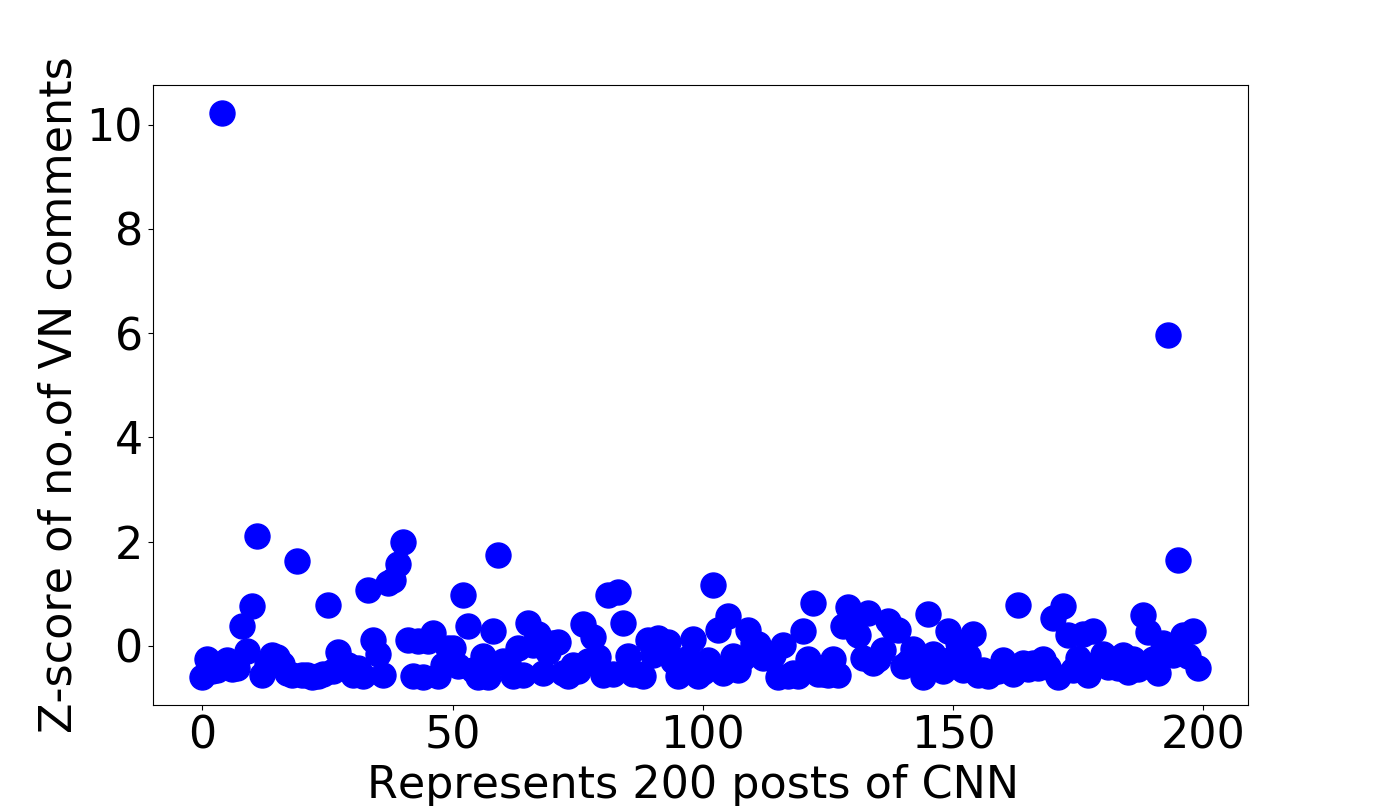}
\label{fig:z-score-cnn}
	}   \hspace{-0.8cm}
	\subfigure[FoxNews]{
  \includegraphics[width=2.4in]{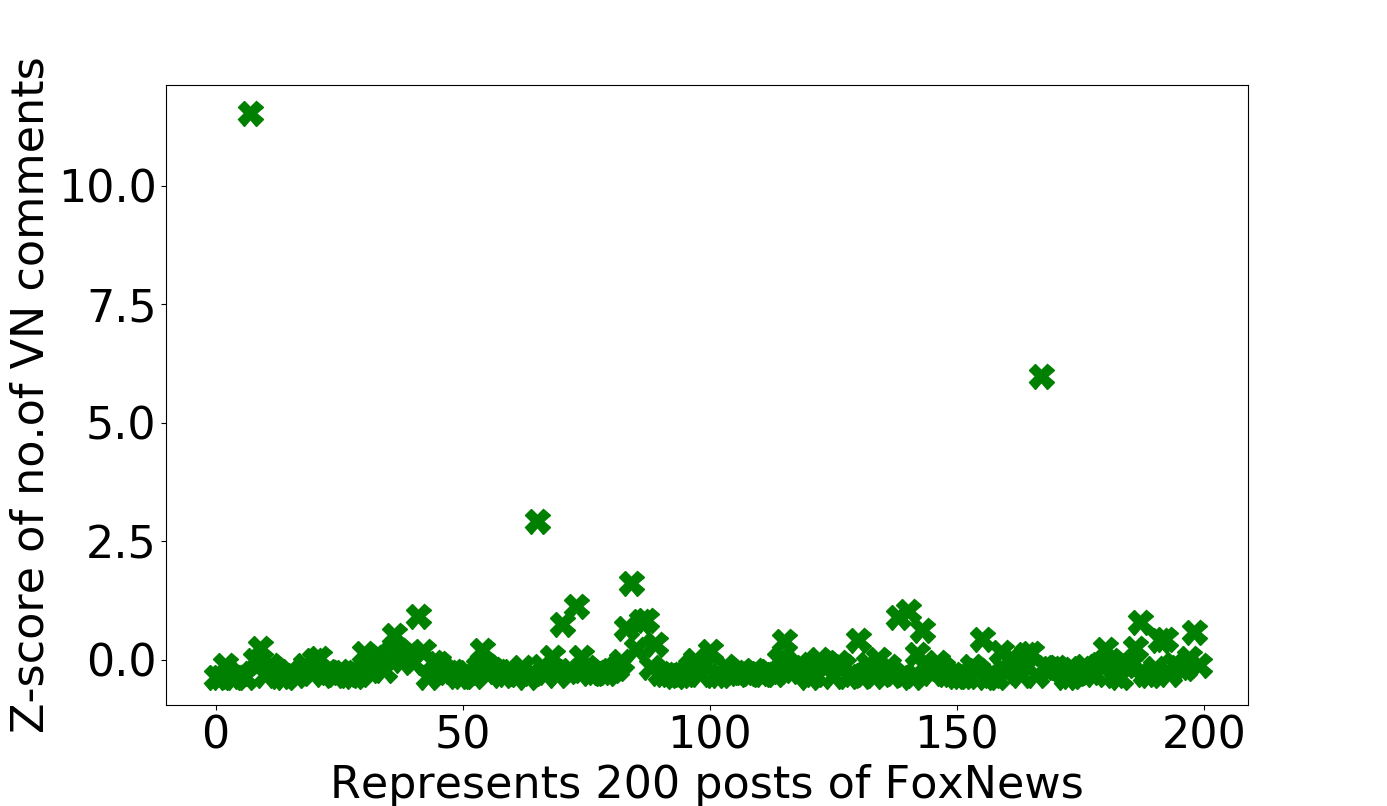}
\label{fig:z-score-fox}
	}
	\caption{The distribution of z-score values of number of Very Negative (VN) comments received per post (200 posts in total).}
		\vspace{-0.1cm}
	\label{fig:z-score}
\end{figure*}

\begin{table*}[]
\centering
\caption{Statistics of the flaming types of posts shared by BBCNews, CNN and FoxNews (VN stands for Very Negative).}
\label{tab:flaming-dataset}
\begin{tabular}{|l|l|l|l|p{10cm}|}
\hline
\multicolumn{1}{|c|}{\textbf{News media}} & \multicolumn{1}{c|}{\textbf{Date}} & \multicolumn{1}{c|}{\textbf{\#Reactions}} & \multicolumn{1}{c|}{\textbf{\#Comments(VN)}} & \multicolumn{1}{c|}{\textbf{Topics}} \\ \hline
\multirow{3}{*}{BBCNews}                  & 2018-02-01                         & 13590                                     & 4110 (44.28\%)                                     & Palestinian teenager slapped an Israeli soldier.                          \\ \cline{2-5} 
                                          & 2018-02-02                         & 33110                                     & 3946 (37.56\%)                                     & A father has tried to attack a doctor who abused his three daughters.                                   \\ \cline{2-5} 
                                          & 2018-02-14                         & 28718                                     & 2614 (63.54\%)                                     & Shooting at a Florida high school.                                    \\ \hline
\multirow{2}{*}{CNN}                      & 2018-02-02                         & 113632                                    & 19598 (30.99\%)                                    & A father has tried to attack a doctor who abused his three daughters.                   \\ \cline{2-5} 
                                          & 2018-02-14                         & 39611                                     & 4670 (73.17\%)                                     & Shooting at a Florida high school. \\ \hline
\multirow{2}{*}{FoxNews}                  & 2018-01-31                         & 86717                                     & 130185 (20.61\%)                                   & House Democratic Leader Nancy Pelosi and other House Democratic leaders hold a news conference.
                                    \\ \cline{2-5} 
                                          & 2018-02-13                         & 72765                                     & 35619 (29.41\%)                                    & The women of 'The View' took a shot a Vice President Mike Pence's Christian faith on Tuesday, mocking the former governor of Indiana for talking to Jesus and even calling it a 'mental illness'.                                    \\ \hline
\end{tabular}
\end{table*}

\subsection{Flaming events in news media - time-series approach}
Receiving avalanches of negative emotions or flood of low happiness comments from users within short period  of time tend to generate flaming events and a time-series approach can be one of the optimal ways to explore it.
Figure \ref{fig:sentimentdistro_negative} exhibits the time-series distribution of received Negative and Very Negative comments in our experimental dataset. Figure \ref{fig:timeseries-perday} exhibits negative comments received per day during February 2018. As shown in Figure \ref{fig:timeseries-perday}, number of Very Negative comments is always higher than the number of Negative comments and therefore we focus only on Very Negative comments for further analysis.
We can observe that, on 14th February 2018, a flood of negative comments have received on posts shared by all three news media compared to the other days. 
Apart from that, 31st January, 2nd and 5th February exhibited another flaming type of distributions showing few other spikes of number of Very Negative comments.

As shown in Figure \ref{fig:timeseries-perday}, the main flaming type of event taken place in 14th February. 
Therefore, to identify more information about this event the consideration of the posts shared during one day before and after 14th February is also important. Figure \ref{fig:timeseries-14thfeb} shows this distribution in terms of the Very Negative comments received during these 3 days. We can observe that all three  news media have shared the flaming posts in 14th February and received a large number of Very Negative comments but, FoxNews has shown a flaming type of event on 13th February where some of the comments on these posts have received on 14th February as well. One main inspection from these figures indicate our hypothesis that the flaming type of posts received Very Negative comments within short duratin of time (approximately 2-3hrs) after posting news.

Figure \ref{fig:timeseries-perday} and \ref{fig:timeseries-14thfeb} indicate varied values of number of comments received on 14th February. 
We explored that on 14th February, BBCNews, CNN, FoxNews have published only 14.5\% posts, 19\% posts and 16.5\% posts respectively out of 200 posts. 
Next we identified how many posts have received a higher number of negative comments than the other types of comments. The statistics shown that only 26.6\% posts from BBCNews, 34.2\% from CNN and 30.3\% from FoxNews have received more negative comments than the other types of comments on 14th February. Hence, as per this analysis, only a set of posts contributed to the flaming event on 14th February 2018 while other comments have received on the posts shared by the other days. Next section will explain more about flaming event posts and their discussed topics based on one statistical approach. 

\subsection{Flaming events in news media - statistical approach}
The analysis of comments based on the time-series data has proven an existence of flaming event in BBCNews, CNN and FoxNews. 
Identification of types of the posts of these flaming events are useful and therefore in this section we will explore more details about the posts that attracts flaming comments. 

At first, we try to identify these flaming posts with the use of one statistical approach that can predict these types of behaviors considering an outliers prediction method where outliers in this scenario are the posts those that received many insults or offensive comments. The standard score (z-score) is one outlier prediction method which allows to identify whether a particular value is equal to the mean, below the mean or above the mean. We calculate z-score for a distribution of number of Very Negative comments received on each posts within February 2018 using following equation where $\mu$ and $\sigma$ represents mean and standard deviation, respectively. For each post we assign 'x', the number of Very Negative comments.
$$z = \frac{x - \mu}{\sigma}$$
Figure \ref{fig:z-score} exhibits the distribution of z-score values of Very Negative comments received on posts in our experimental dataset. As shown in Figure \ref{fig:z-score-bbc}, BBCNews has three outliers having z-score value greater than 5. Figure \ref{fig:z-score-cnn} exhibits two outliers for CNN with the z-score above 6 and as shown in Figure \ref{fig:z-score-fox}, FoxNews has two outliers having z-score greater than 6. Next, we will analyze the popularity of these 7 outliers posts and their discussed topics. 

Table \ref{tab:flaming-dataset} contains the information on above identified 7 outliers including published date, total number of reactions received, number of comments received, number of Very Negative comments (VN) and the discussed topics. 
We observed that, these posts are widely popular and have received enormous amount of reader reactions (\#reactions) compared to the other posts in the dataset. We also observed that these posts have received huge amount of comments and more of them are Very Negative (\#comments). In addition, all 7 posts have received more than 20\% of the comments as Very Negative.

An interesting observation of this analysis is to explore the types of topics discussed by each news media in these outlier posts. Table \ref{tab:flaming-dataset} exhibits that BBCNews and CNN have posted similar topics on i) 2nd February - about a father was trying to kill a doctor who abused his daughters and ii) 14th February - about an incident of shooting at a Florida high school. In addition, BBCNews has an extra flaming event on iii) 1st February - about one Palestinian teenager slapping an Israeli soldier. 
On the other hand, two flaming posts published by FoxNews are different to what BBCNews and CNN have shared; on iv) 31st January - type of a political discussion and v) 13th February -  related to a religious discussion.
In summary, we identified only 5 posts associated to the 7 outliers detected in the previous section. Therefore, we can conclude that, as exhibited in Figure \ref{fig:sentimentdistro_negative} 31st January and 2nd, 13th, 14th February exhibited flaming types of news posts with the analysis of a statistical method. The topics discussed on these flaming posts are mainly very sensitive information, political related details or related to religious matters. We can deduce from these information that, people try to send more aggressive and negative comments on sensitive news items.

These results conclude that we can detect flaming events in news media in Facebook using a sentiment prediction approach, and we can apply this analysis to other types of categories in Facebook such as celebrities, politicians, etc. Moreover, if the SCL-NMA dataset can be improved by adding more hateful and insult wordings with their sentiment scores, we can detect flaming events with higher accuracy. 
Our deep NN model can also be used to adapt to other use-cases, as it is a general model to detect the sentiments of textual content from five different classes. The proposed model can be enhanced to identify flaming wars in SNSs as the model we developed can be automated and behave as a semi-supervised approach. The comments received on flaming posts might be shared by real users or might be Spam generated content \cite{44}. Therefore, it is important to identify these patterns of posting flaming events based on commentator's behavioral properties.



%% file: 6_Conclusion.tex
\section{Conclusion}
\label{lbl:conclusion}
This paper is mainly focused on the detection of flaming types of events in news media in Facebook.
We proposed a deep Neural Network (NN) model to predict sentiment polarity of SNSs comments based on word embedding. We also proposed an improved lexicon-based multiclass sentiment classification method to generate true labels for our NN model. We experimented with two word embedding models: Word2Vec and FastText and found that Word2Vec model performed better (85\% accuracy) than FastText model (78\%). 
The model returns five sentiment classes; very negative, negative, neutral, positive and very positive. 
As a use-case, we explored some existing flaming types of events in BBCNews, CNN and FoxNews Facebook pages. A flaming event can be present in SNSs when a post receives many negative comments withing a short period of time. However, we identified that, these news media received a higher number of neutral comments than other categories.
First, we explored flaming behavior based on a time-series analysis of the comments received on posts during February 2018 and a major flaming event is identified on 14th February 2018. Then, using one statistical based approach we justified the existence of flaming types of incidents in news items that were identified from the time-series analysis. The results shown that, in news media, flaming types of events are more common when they published very sensitive information (murders, rape etc.), political based contents and information related to religious belief.
  